\documentclass{amsproc}
\begin {document}

\title{Gauge theories on noncommutative spaces}

\author{Albert Schwarz}

\address{Department of Mathematics, University of California at Davis,
Davis, CA 95616}

\email{schwarz@math.ucdavis.edu}



\begin{abstract}
   I review my results  about noncommutative gauge theories and about the
relation of these theories to  M(atrix)  theory following my lecture on
ICMP 2000.
\end{abstract}

\maketitle

In my lecture on ICMP 2000  I gave a short review of my results  on
noncommutative gauge theories and talked in more detail about my recent
paper [9]. Here I'll  skip all details referring to papers [1]-[13]. I'll
list only main results of these papers. 

In the paper [1] it was shown  that gauge theories  on  noncommutative
tori appear naturally in consideration of compactifications of M(atrix)
theory. The same logic can be used to obtain gauge theories on
noncommutative toroidal orbifolds [14],  [15], [11], [12]. 

More precisely, if  $G$ is a subgroup of the group of symmetries of any
model we can restrict ourselves to fields that are $G$-invariant up to
gauge equivalence. This means that the change of a field $A$ under the
action of an element $\gamma\in G$ can be compensated for by gauge
transformation $U_{\gamma}$.  For matrix models (i.e. in the case when $A$
is a collection of matrices and gauge transformations are unitary
transformations) this means that 
\begin {equation}
\gamma (A)=U_{\gamma}AU_{\gamma}^{-1}.
\end {equation}
Usually finite size matrices don't satisfy this equation; one should
replace (Hermitian) matrices by 
(Hermitian) operators in infinite-dimensional Hilbert space $E$ and
consider $U_{\gamma}$ as unitary operators in this space. There exists  no
reason to expect that $U_{\gamma \lambda}=U_{\gamma}U_{\lambda}$, but
taking into account that 

\begin {equation}
(U_{\gamma\lambda}^{-1}\cdot
U_{\gamma}U_{\lambda})A(U_{\gamma\lambda}^{-1}\cdot
U_{\gamma}U_{\lambda})=A
\end {equation}
it is naturally to assume that 

\begin {equation}
U_{\gamma\lambda}=e^{i\pi\theta(\gamma ,\lambda) }U_{\gamma}U_{\lambda}
\end {equation}

One can say, that  the operators $U_{\gamma}$ specify a projective
representation of the group  $G$. In
the case when $G={\bf Z}^d$ the associative algebra $T_{\theta}^d$
generated by operators $U_{\gamma}$ can be interpreted as the algebra of
functions  on $d$-dimensional noncommutative torus. In other words the
space $E$ can be considered as a $T_{\theta}^d$-module. We always consider
finitely generated projective modules (direct summands in free modules
$(T_{\theta}^d)^n$).      In  noncommutative geometry this means   that we
consider  "vector bundles" over  noncommutative tori. 

The torus
$T_{\theta}$ is specified by means of bilinear form $\theta (\gamma
,\lambda)$ on ${\bf Z}^d$; without loss of generality one can assume that
this form is antisymmetric. It will be more convenient for us to say that
a  noncommutative torus is determined by antisymmetric matrix
$\theta_{jk}$  corresponding to the form $\theta (\gamma ,\lambda)$ in
some basis of ${\bf Z}^d$. In terms of this matrix   noncommutative torus
can be interpreted as an algebra with unitary generators $U_1,...,U_d$
satisfying relations $U_jU_k=e^{2\pi i\theta_{jk}}U_kU_j$.
If $A=(A_1,...,A_d)$ and the group ${\bf Z}^d$ acts on $A$ by means of
translations (i.e. $\gamma(A)=A+\gamma$), then the solution of the
equation (1) can be considered as a connection on     noncommutative torus
$T_{\theta}$ in the sense of A.Connes [16], [17]. (The notion of
connection is discussed in detail at the end of the paper.)  If our
starting point is BFSS or IKKT matrix model [18], [19],then the above
construction leads to SUSY Yang-Mills theory on  noncommutative torus [1].
Replacing ${\bf Z}^d$ with semidirect product of ${\bf Z}^d$ and finite
group we obtain gauge theories on  noncommutative toroidal orbifolds. 
The appearance of   noncommutative geometry can be explained  not only
from the viewpoint of M(atrix) theory, but also from the viewpoint of
string theory as was shown in a series of papers [20]-[24], culminating by
Seiberg-Witten paper [25] that contains very detailed analysis of relation
between string theory and gauge theory on  noncommutative spaces. 

Gauge theories on    noncommutative tori were studied by A.Connes and M.
Rieffel, especially in two-dimensional case [26]-[28]. I obtained new
results  about these theories  focusing my attention on problems related
to physics. Already in [1] it was conjectured that Morita equivalence  of
algebras is related to duality in physics. One says that an algebra $A$ is
Morita equivalent to the algebra $\hat {A}$ if the category of $A$-modules
is   equivalent to the category of $\hat {A}$-modules. In other words, we
should be able to transfer $A$-modules into $\hat {A}$-modules and $\hat
{A}$-modules into $A$-modules;  this correspondence should be natural (for
every $A$-linear map $\varphi :E\rightarrow E^{\prime}$ of $A$-modules
should be defined an $\hat {A}$-linear map $\hat {\varphi} :\hat
{E}\rightarrow \hat {E}^{\prime}$  of corresponding $\hat {A}$-modules;
one requires that the correspondence ${\varphi} \rightarrow \hat
{\varphi}$ transforms composition  of maps
into composition of maps ). However, to prove that gauge theories over $A$
are related to gauge theories on $\hat {A}$ we should be able to transfer
also connections on $A$-modules to connections on $\hat {A}$-modules. I
introduced a new notion of gauge  Morita equivalence (in original paper
[2] I used the term "complete  Morita equivalence") and proved that gauge
Morita equivalence of algebras implies physical equivalence of
corresponding gauge theories.It is proved in [2] that   noncommutative
tori $T_{\theta}^d$ and $T_{\hat{\theta}}^d$ are gauge  Morita equivalent
if and only if there exists a matrix  
\begin{equation}
\left( \begin{array}{cc}
A & B \\
C & D
\end{array} \right)
\end{equation}
belonging to $SO(d,d,{\bf Z})$ 
and obeying 
\begin {equation} 
\hat {\theta}=(A\theta +B)(C\theta +D)^{-1}
\end {equation}
Here $A, B, C, D$ are $d\times d$ matrices and 
$SO(d,d,{\bf Z})$ 
stands
for the group of $2d\times 2d$ matrices with integer entries that are
orthogonal with respect to quadratie form $x_1x_{d+1}+...+x_dx_{2d}$
having signature $(d,d)$. (The fact that the relation (5) implies  Morita
equivalence was proved in earlier paper [4] written together with M.
Rieffel.) Equivalence of gauge theories on noncommutative tori
$T_{\theta}$ and $T_{\hat {\theta}}$ was studied in detail in [2], [6],
[7], [8]. It is closely related to $T$-duality in string theory; this
relation was thoroughly analyzed in [5]. This analysis led, in particular,
to the discovery of possibility to trade noncommutativity parameter for
background field in the expressions for BPS energies. (Almost
simultaneously this fact  was found in [25] at the level of action
functionals; it was called background independence.)  

The papers [6], [7], [8], [13]    are devoted to the study of BPS fields
and BPS states in SUSY gauge theories on noncommutative tori. Analysis of
BPS spectra by means of supersymmetry algebra was performed in [7].
Another way to study BPS states is based on geometric quantization of
moduli spaces of classical configurations having some supersymmetry (BPS
fields). One can identify ${1\over 2} $ BPS fields with connections having
constant curvature. We found necessary and sufficient for existence of
constant curvature connections and described moduli spaces of such
connentions. As in the case of commutative torus the $K$-theory class of a
projective module $E$ (of a "vector bundle") over $T_{\theta}^d$ can be
characterized by means of an even integer element $\mu(E)$ of Grassman
algebra with $d$ generators $\alpha ^1,...,\alpha ^d$ (as a collection of
integer antisymmetric tensors $\mu ^{(0)},\mu
_{ij}^{(2)},\mu_{ijkl}^{(4)},....$ of even rank). One can interpret
$\mu^{(0)},\mu_{ij}^{(2)},...$ as numbers of D-branes with given
topological charges (although a notion of individual D-brane is
ill-defined in noncommutative  space). We will assume that $\theta$ is
irrational; then it was proved in [27] that two modules having the same
$K$-theory class are isomorphic. One can verify that  necessary and
sufficient condition for existence of constant curvature connection in a
module with $\mu(E)=\mu$ is a possibility to represent $\mu$ as a
quadratic exponent (as an expression of the form $C exp(\alpha
^k\rho_{kl}\alpha ^l)$) or as a limit of quadratic exponents. If a module
$E$ having  constant curvature connection cannot be represented as a
direct sum of isomorphic modules (i.e. g.c.d.$(\mu ^{(0)},\mu
_{ij}^{(2)},\mu_{ijkl}^{(4)},....)=1)$ we say that $E$ is a basic module.
For basic $T_{\theta}^d$-module the moduli space of constant curvature
connections  is a $d$-dimensional torus $T^d$. An arbitrary
$T_{\theta}^d$-module with constant curvature connection is isomorphic to
a direct sum of $n$ basic modules where $n=$g.c.d.$(\mu ^{(0)},\mu
_{ij}^{(2)},\mu_{ijkl}^{(4)},....)$; for such a module corresponding
moduli space is $(T_{\theta}^d)^n/S_n$.
Basic modules and constant curvature connections on these modules can be
described very explicitly  in the language of  so called Heisenberg
modules.
Every $T_{\theta}^d$-module can be represented as a direct sum of basic
modules, the number  of summands in this sum can be made as large as we
want. This statement follows from results of [27]. (Recall that we assumed
irrationality of $\theta$; this assumption is  necessary and sufficient
for validity of our claim.) This means, in particular, that every
combination of  D-branes can decay into ${1\over 2}$BPS states.
 
Instantons on noncommutative ${\bf R}^4$ were analyzed in [3] by means of
generalization of ADHM construction. The most striking feature of
noncommutative instantons is the absence of small instanton singularity in
moduli space of noncommutative instantons. Instantons on noncommutative
tori were studied in [10]; in particular, we constructed a noncommutative
analog of Nahm transform. Instantons can be characterized as ${1\over
4}$BPS fields. In the case when a $T_{\theta}^d$-module admits    constant
curvature connection it is possible to give a complete description of
${1\over 4}$BPS fields and ${1\over 4}$BPS states [6]. To obtain this
description  one can apply the fact that under certain conditions on
$\theta$ we can use Morita equivalence to transform such a module into a
free module over another noncommutative torus. (It is sufficient to assume
that every linear combination of matrix elements of $\theta$ having
integer coefficients is irrational.) This remark can be used also in many
other cases; it confirms the idea that noncommutative tori with irrational
$\theta$ are simpler that commutative tori.

Gauge theories on noncommutative toroidal orbifolds were studied in [11],
[12]. Fairly complete analysis of modules, of constant curvature
connections and corresponding moduli spaces, of Morita equivalence is
given  for $T_{\theta}^d/{\bf Z}_2$; however, the methods developed in
[11], [12]  work also for other  toroidal orbifolds. 

All results we mentioned are based on the notion of connection on
$A$-module. There  exist different definitions of this notion, but all of
them are based on the same idea: a connection should satisfy Leibniz
rule. If an $n$-dimensional Lie algebra $L$ acts on associative algebra
$A$ by means of infinitesimal automorphisms (derivations) we can define a
connection on (left) $A$-module $E$ as a collection of $n$ linear
operators $\nabla_i:E\rightarrow E,\ \  i=1,...,n$ obeying the Leibniz
rule:
$$\nabla_i(ae)=a\cdot \nabla_ie+\delta_ia\cdot e,$$
where $a\in A,\ \  e\in E$, and $\delta_1,...,\delta_n$ are derivations
corresponding to elements of a basis of Lie algebra $L$. (Notice, that
operators $\nabla_i$ don't commute with multiplication by $a\in A$, i.e.
they are ${\bf C}$-linear, but not $A$-linear. However, if $\nabla_i\ ,
\nabla_i^{\prime}$ ($i=1,...,n$) are two connections the difference
$\nabla_i^{\prime}-\nabla_i$ is $A$-linear ; in other words
$\nabla_i^{\prime}-\nabla_i$ is an endomorphism of $E$.) 

When we consider  noncommutative tori we should define connections  using
$d$-dimensional commutative Lie algebra acting on $T_{\theta}^d$ by means
of translations.

If we would like to define connections in terms of covariant differential
instead of covariant derivative we should assume that the algebra $A$ is a
${\bf Z}_2$-graded associative algebra equipped with a parity reversing
derivation $Q:A\rightarrow A$. The standard assumption is that $Q^2=0$
(then $A$ is called a graded differential algebra). However it is shown in
[9] that one can relax this assumption requiring only that
$Q^2a=[\omega,a]$. (Here $\omega$ is a fixed element of $A$ obeying
$Q\omega =0$.) If $A$ is an associative algebra equipped with an operator
$Q$ of this kind (a $Q$-algebra is the terminology of [9]) we can  define
a connection on (left) $A$-module $E$ as  an linear operator $\nabla
:E\rightarrow E$ obeying the Leibniz rule:
$$D(ae)=(-1)^{deg a}aDe+Qa\cdot e.$$ 
The standard theory of  connections (including the notion of Chern
character) can be generalized to the case of  modules over a $Q$-algebra.
If $P$ is a  module over $Q$-algebra $A$ and $\nabla_P$ is a connection on
$P$ we can define a structure of $Q$-algebra on $\hat {A}={\rm End}_AP$ by
the formula $\tilde {Q}\varphi=[\nabla_P,\varphi]$. (Here ${\rm End}_AP$
stands for an algebra of endomorphisms of $A$-module $P$, i.e. for an
algebra of $A$-linear maps of $P$ into itself.) Under certain conditions
on $P$ 
 the algebra $\hat {A}$ is
Morita equivalent to $A$, i.e. we can transfer  $A$-modules into
$\hat{A}$-modules and vice versa.
(If $A$ has a unit we should require that $A$ considered as left
$A$-module is a direct summand
of $P^N$ for some $N$ and $P$ is projective)
Using the connection $\nabla_P$ we can transfer  connections on
$A$-modules  to  connections on corresponding on
$\hat{A}$-modules. This operation permits us to extend the  equivalence
between categories of $A$-modules and $\hat{A}$-modules to an 
equivalence of corresponding 
gauge theories . This gives a very
general duality theorem; $SO(d,d,{\bf Z})$ duality of gauge theories on
noncommutative tori can be derived from this general theorem [9].

\bigskip
\centerline {\bf References.}

  1.  A. Connes,  M. Douglas,  and  A. Schwarz,  {\it  Noncommutative
Geometry and Matrix Theory: Compactification on  Tori},  JHEP {\bf 02}
(1998),  3-38

   2. A.  Schwarz, {\it  Morita Equivalence and Duality},  Nucl. Phys.
{\bf  B 534} (1998), 720-738.

   3. N. Nekrasov  and  A. Schwarz,  {\it Instantons on Noncommutative
$R^4$
and (2,0) Superconformal Six-dimensional Theory}, Comm. Math. Phys.{\bf
198} (1998), 689-703.

   4. M. Rieffel  and  A. Schwarz, {\it Morita Equivalence of
Multidimensional Noncommutative Tori},  Intl. J. of Math  {\bf 10(2)}
(1999), 289-299

   5. B. Pioline and A. Schwarz, {\it Morita Equivalence and T-Duality},
JHEP {\bf 9(21)} (1999),  1-16

   6. A. Konechny  and  A. Schwarz,   {\it 1/4 - BPS States on
Noncommutative Tori}, JHEP. {\bf  30} (1999), 1-14 
 
   7. A. Konechny and  A. Schwarz,  {\it Supersymmetry Algebra and BPS
States of super Yang-Mills Theories on Noncommutative Tori},  Phys. Lett.
{\bf B 453} (1999),  23-29

   8. A. Konechny  and  A.  Schwarz, {\it  BPS States on Noncommutative
Tori and Duality}, Nucl. Phys. {\bf B 550}  (1999), 561-584.

   9. A. Schwarz, {\it Noncommutative Supergeometry and Duality},  Lett.
Math. Phys. {\bf  50 (4)} (1999),  309-321

  10. A. Astashkevich,  N. Nekrasov   and  A. Schwarz,  {\it On
Noncommutative Nahm Transform}, Comm. Math. Phys. {\bf  211 (1)} (2000),
167-182.

  11. A. Konechny  and  A. Schwarz,  {\it  Moduli Spaces of Maximally
Supersymmetric Solutions on Noncommutative Tori and  Noncommutative
Orbifolds},  JHEP {\bf 09} (2000), 1-23. 

   12. A. Konechny  and  A. Schwarz,  {\it  Compactification of M(atrix)
Theory on Noncommutative Toroidal Orbifolds},  Nucl. Phys. {\bf  B 591(3)}
(2000), 667-684.

   13. A. Astashkevich and  A. Schwarz,  {\it  Projective Modules Over
Noncommutative Tori: Classification of Modules with Constant Curvature
Connection}, Journal of Operator Theory (in press).

  14. M. Douglas, {\it D-branes and Discrete Torsion},  hep-th/9807235.
   
  15. P.M. Ho, Y.Y. Wu and Y.S. Wu, {\it Towards  Noncommutative Geometric
Approach to  Matrix Compactification}, Phys.Rev. {\bf  D 58} (1998), 26006. 

 16. A. Connes, {\it  C*-algebres et Geometrie Differentielle}, C. R.
Acad. Sci. Paris {\bf 290} (1980), 599-604.

17. A. Connes, {\it Noncommutative Geometry}, Academic Press, 661 pp. 

18. T. Banks, W. Fischler, Shenker, and I.Susskind, {\it M-theory as a
Matrix Model: a Conjecture},  Phys. Rev. {\bf D 55} (1997), 5112.

  19. N. Ishibashi, H. Kawai, I. Kitazawa, and Tsuchiya, {\it A Large-N
educed Model as Superstring}, Nucl. Phys. {\bf B 492} (1997),  467-491.

  20. M. Douglas and C. Hull, {\it D-branes and  Noncommutative Torus},
JHEP {\bf 02} (1998), 008

 21. Y.-K. E. Cheung and M. Krogh, {\it Noncommutative Geometry from
0-Branes in a Background B Field}, Nucl. Phys. {\bf B 528} (1998), 185.

 22. C.-S. Chu and P.-M. Ho, {\it Noncommutative Open String and D-Brane},
Nucl. Phys. {\bf B550} (1999), 151;  {\it Constrained Quantization of Open
String  in  Background B Field and Noncommutative D-Brane},  Nucl. Phys.
{\bf B 568} (2000),  447. 

23. V. Schomerus, {\it D-Branes and Deformation Quantization }, JHEP {\bf
06} (1999), 030.

24. F. Ardalan, H. Arfaei and M. Sheikh-Jabbari, {\it Mixed Branes and
M(atrix) Theory on Noncommutative Torus},  PASCOS 98; {\it  Noncommutative
Geometry from String and Branes}, JHEP {\bf 02} (1999), 016; {\it Dirac
Quantization of Open Strings and   Noncommutativity in Branes},
Nucl.Phys. {\bf B 576} (2000), 578-596.

  25. N. Seiberg and E. Witten, {\it String Theory and Noncommutative
Geometry}, JHEP {\bf 09} (1999),  032.

  26. A. Connes and  M. Rieffel, {\it Yang-Mills for  Noncommutative
Two-Tori}, Contemporary Math. {\bf 66} (1987), 237-266.

   27. M. Rieffel, {\it Projective Modules over Higher-dimensional
Noncommutative Tori}, Can. J. Math., Vol. {\bf XL}, No. 2 (1998), 257-338.

  28.  M. Rieffel, {\it  Critical Points of Yang-Mills for  Noncommutative
Two-Tori},  J. Diff. Geom., 
{\bf  31} (1990), 535.

\end{document}